\documentclass[12pt]{article}
\usepackage{titling}
\usepackage{amsmath}
\usepackage{slashed}
\usepackage{amssymb}
\usepackage{epsfig}
\usepackage{graphicx}
\usepackage{placeins}
\usepackage{multirow}
\usepackage{color}
\usepackage[normal]{subfigure}
\usepackage{rotating}
\usepackage{hyperref}
\usepackage[margin=0.9in]{geometry}
\usepackage[table]{xcolor}
\usepackage{enumitem}
\usepackage[utf8x]{inputenc}
\usepackage[compress,numbers,sort]{natbib}
\usepackage{authblk}
\usepackage{colortbl}
\usepackage{pdflscape}
\usepackage{color}
\usepackage{mathtools}
\usepackage{tabu}
\usepackage{braket}
\usepackage{arydshln}

\newcommand{\published}[1]{%
\gdef\puB{#1}}
\newcommand{\puB}{}

\title{Revisiting R-parity violating interactions as an explanation of the B-physics anomalies}

\author[1]{\large Sokratis Trifinopoulos\thanks{trifinos@physik.uzh.ch}}
\affil[1]{\emph{\normalsize Physik-Institut, Universit\"at Z\"urich, CH-8057 Z\"urich, Switzerland}}

\date{}

\published{
\begin{flushright}

\end{flushright}
\vskip 2cm }

\begin{document} 

\maketitle

\begin{abstract}
\noindent
In the last few years, the ratios $R_{D^{(*)}}$ and of $R_{K^{(*)}}$ have reportedly exhibited significant deviations from the relevant Standard Model predictions, hinting towards  a possible violation of Lepton Flavor Universality and a window to New Physics. We investigate to what extent the inclusion of R-parity violating couplings in the Minimal Supersymemtric Standard Model can provide a better fit to the anomalies simultaneously. We perform this analysis employing an approximate, non-abelian $\mathcal{G}_f=U(2)_q \times U(2)_\ell$ flavour symmetry, which features a natural explanation of the appropriate hierarchy of the R-parity violating couplings. We show that, under the requirement of a supersymmetric spectrum with much heavier left-handed doublet superpartners, our assumption favors a considerable enhancement in the tree-level charged-current $B \to D^{(*)} \tau \overline{\nu}$, while the anomalies induced by $b \to s\ell^+\ell^-$ receive up to an approximate $30 \%$ improvement. The consistency with all relevant low-energy constraints is assessed.
\end{abstract}

\clearpage
\tableofcontents
\clearpage

\section{Introduction}

While exploring the limits of the Standard Model (SM), many experiments have spent decades looking for processes that do not respect Lepton Flavor Universality (LFU). Until 2014 this fundamental SM feature stood rather steady, but since then several B-physics experiments, started reporting results which conflict with it, to the surprise and excitement of the community. These results are encoded by ratios of branching ratios involving rare $b$ decays and different lepton flavors. 
\begin{enumerate}
	\item For the case of the \textit{charged-current} interactions, we define:
	\begin{align}
R_{D^*} &= \frac{ \mathcal B(B \to D^* \tau \overline{\nu})}{ \mathcal B(B \to D^* \ell \overline{\nu} ) }, \label{eq:RDS}  \\
R_D &= \frac{ \mathcal B(B \to D  \tau \overline{\nu}) }{ \mathcal B(B \to D  \ell \overline{\nu} ) }, \label{eq:RD}
\end{align}
where $\ell=e,\mu$ for BaBar and Belle, while $\ell=\mu$ for LHCb. \par
The experimental world averages of Babar~\cite{Lees:2013uzd}, Belle~\cite{Hirose:2016wfn}, and LHCb data~\cite{Aaij:2015yra} from the Heavy Flavor Averaging Group~\cite{Amhis:2016xyh} read
	\begin{equation}
R^{\text{exp}}_{D^*} = 0.310 \pm 0.015 \pm 0.008, \ \ \ \ \ \ \ R^{\text{exp}}_{D} = 0.403 \pm 0.040 \pm 0.024,
	\end{equation}
to be compared with the theory predictions~\cite{Bernlochner:2017jka}~\cite{Bigi:2017jbd}
	\begin{equation}
R^{\text{SM}}_{D^*} = 0.257 \pm 0.003, \ \ \ \ \ \ \ R^{\text{SM}}_D = 0.299 \pm 0.003. \label{eq:RDSexp}
\end{equation}

\item For the case of the \textit{flavor-changing neutral-current (FCNC)}  interactions, we define:
\begin{align}
R_{K^*} &= \frac{ \mathcal B(B \to K^* \mu\overline{\mu}) }{ \mathcal B(B \to K^* e \bar{e} )}, \label{eq:RK}  \\
R_K &= \frac{ \mathcal B(B \to K \mu\overline{\mu}) }{ \mathcal B(B \to K e \bar{e} )}. \label{eq:RKS}
\end{align}
The LHCb Collaboration measured these rations in the di-lepton invariant mass bin \\ $q^2\in[1,6]{~\rm GeV}$ and found ~\cite{Aaij:2014ora}~\cite{Aaij:2017vbb}
	\begin{equation}
\left. R^{\text{exp}}_{K^*} \right|_{q^2\in[1.1,6]{~\rm GeV}} =  0.685^{+0.113}_{-0.069} \pm 0.047, \ \ \ \ \
R^{\text{exp}}_K = 0.745^{+0.090}_{-0.074} \pm 0.036.
	\end{equation}
The SM expectation value are with percent level accuracy~\cite{Bordone:2016gaq}
	\begin{equation}
\left. R^{\text{SM}}_{K^*} \right|_{q^2\in[1.1,6]{~\rm GeV}} = R^{\text{SM}}_{K} = 1.00 \pm 0.01.
	\end{equation}
In the lower $q^2\in[0.045,1.1]{~\rm GeV}$ bin, the experimental value for $R_{K^*}$~\cite{Aaij:2017vbb} is 
	\begin{equation}
\left. R^{\text{exp}}_{K^*} \right|_{q^2\in[0.045,1.1]{~\rm GeV}} =  0.660^{+0.110}_{-0.070} \pm 0.024,
	\end{equation}
but the theory prediction is more delicate due to threshold effects and implies larger theoretical uncertainities~\cite{Bordone:2016gaq}
	\begin{equation}
\left. R^{\text{SM}}_{K^*} \right|_{q^2\in[0.045,1.1]{~\rm GeV}} =  0.906 \pm 0.028.
	\end{equation}	
\end{enumerate}
The statistical significance of each anomaly does not exceed the $3 \sigma$ level, but the overall set is very consistent and the pattern of deviations intriguing. Our analysis will be solely
focused on the above mentioned LFU ratios; nevertheless, it is worth mentioning briefly the existence of additional data that exhibit tensions with the SM predictions. In particular, the most notable is a deviation of about $3\sigma$ reported~\cite{Aaij:2015oid}~\cite{Wehle:2016yoi} on the so-called $P_5^\prime$ differential observable of $B\to K^*\mu\overline{\mu}$ decays. Even though, given the non-negligible SM uncertainties~\cite{Ciuchini:2015qxb}, the $P_5^\prime$ anomaly is also not an individually definite NP signal, it has been pointed out that it has a common model-independent solution with the $R_{K^{(*)}}$ anomaly, namely the modification of a single amplitude, the one induced by the semi-leptonic di-muon vector and axial operators ~\cite{Descotes-Genon:2015uva, Altmannshofer:2015sma, Hurth:2016fbr}. \par
While the charged-current decays occur at tree-level, the FCNC decays appear at loop level in the SM, rendering a simultaneous explanation of both anomalies a notoriously difficult theoretical endeavor even in the most general Effective Field Theory (EFT) scenarios without some degree of fine-tuning (see e.g.~\cite{Hiller:2003js, Bobeth:2007dw, Alonso:2015sja, Calibbi:2015kma, Bhattacharya:2015ida, Alonso:2016gym, Nandi:2016wlp, Ivanov:2016qtw, Ligeti:2016npd, Bardhan:2016uhr, Bhattacharya:2016zcw, Alonso:2016oyd, Celis:2016azn, Bhattacharya:2018kig} for model-independent studies and ~\cite{Bauer:2015knc, Hati:2015awg, Fajfer:2015ycq, Cline:2015lqp, Boucenna:2016wpr, Das:2016vkr, Li:2016vvp, Boucenna:2016qad, Becirevic:2016yqi, Sahoo:2016pet, Hiller:2016kry, Bhattacharya:2016mcc, Wang:2016ggf, Crivellin:2016ejn, Popov:2016fzr, Wei:2017ago, Cvetic:2017gkt, Ko:2017lzd, Chen:2017hir, Chen:2017eby, Megias:2017ove, Crivellin:2017zlb, Cai:2017wry, Alok:2017jaf, Alok:2017sui,  Assad:2017iib, DiLuzio:2017vat, Calibbi:2017qbu, Becirevic:2018afm, Marzocca:2018wcf, Blanke:2018sro, Biswas:2014gga, Zhu:2016xdg, Deshpand:2016cpw, Das:2017kfo, Altmannshofer:2017poe, Earl:2018snx, Greljo:2015mma, Barbieri:2015yvd, Bordone:2017anc, Buttazzo:2017ixm, Barbieri:2016las, Bordone:2017bld, Hati:2018fzc} for attempts to cast specific NP models). Moreover, one observes, that the charged-current anomalies require an enhancement in the decay channel that involves the third generation SM fermions, i.e. $b$ and $\tau$, while, as already mentioned, the $b \to s \ell\bar{\ell}$ anomalies are resolved by assuming purely muonic NP effects. \par
The above motivate NP scenarios in which the third generation SM fermions is to be treated specially. On the one hand, the special role of the third generation in the radiative corrections of the Higgs boson self-energy and thus the famous problem of naturalness, is evocative of theories that have traditionally addressed this problem, such as Supersymmetry (SUSY)\footnote{Since direct searches at LHC has been unfruitful so far~\cite{Autermann:2016les}, a `vanilla' MSSM scenario with no fine-tuning is completely ruled out~\cite{Strumia:2011dv}. Extra structure beyond the MSSM field content could prove to be viable model building direction~\cite{Hardy:2013ywa, Arvanitaki:2013yja, Buckley:2016kvr}.}. However, it can be easily checked~\cite{Altmannshofer:2013foa} that the R-parity conserving Minimal Supersymmetric Standard Model (MSSM) introduces amplitudes which are orders of magnitude smaller than the ones required to accommodate the one-loop level anomalies, let alone the tree-level ones. The study of the phenomenology of R-parity violating (RPV) interactions has shown instead that either the charged-current or the FCNC anomalies can be individually resolved~\cite{Biswas:2014gga, Zhu:2016xdg, Deshpand:2016cpw, Das:2017kfo, Altmannshofer:2017poe}, but a united solution which accomodates all other relevant low-energy bounds is impossible. \par
On the other hand, from a flavour point of view, a non-Abelian, $U(2)_q \times U(2)_\ell$ flavour symmetry, acting on the light generations of SM fermions, is one of the most interesting proposals~\cite{Barbieri:2011ci}.  Complemented with the dynamical assumption, that the NP sector is coupled preferentially to third generation, this setup can give a consistent picture for all low-energy flavour observables not only at Effective Field Theroy (EFT) level~\cite{Greljo:2015mma, Barbieri:2015yvd, Bordone:2017anc, Buttazzo:2017ixm} but also in UV complete models, in which $U(2)_q \times U(2)_\ell$ flavour symmetry appears as a subgroup of a greater gauge sector and emerges at low energies ~\cite{Barbieri:2016las}~\cite{Bordone:2017bld}. Interestingly enough, the $U(2)$ symmetries were initially proposed in the context of Supersymmetry~\cite{Barbieri:2011ci}~\cite{Barbieri:1995uv} in order to solve the `flavour' problem of the MSSM, i.e. the abundance of new parameters introduced at the explicit soft SUSY breaking sector and their conspicuous `near-CKM' alignment that avoids unacceptably large flavor-changing and CP-violating effects. \par
By invoking an appropriate flavour symmetry, it is possible to link the RPV sector to the origin of masses and mixings, while naturally suppressing the RPV couplings within the experimental bounds~\cite{Bhattacharyya:1998vw}. In the current work, we employ the $U(2)_q \times U(2)_\ell$ flavour symmetry to give a natural justification to the phenomenologically preferable hierarchy of the RPV couplings. Unlike all previous studies in this framework, we have also taken into account the leptonic current that can be generated by the RPV interactions, besides the usual leptoquark-like current, and how it could affect the relevant amplitudes. A final fit within the natural region of the parameter space reveals certain implications for the SUSY spectrum. \par
The paper is organised as follows. In Section \ref{sec:RPV_and_U2} we review the RPV sector and the $U(2)_q \times U(2)_\ell$ flavour symmetry. We suggest a suitable symmetry breaking pattern and rewrite the RPV couplings with the help of the resulting spurions. Subsequently, in Section \ref{sec:constraints} we examine the relevant latest, low-energy constraints and in Section \ref{sec:fit} we perform a $\chi^2$-fit of the anomalies including those constraints. Finally, the discussion of the results is summarized in the Conclusions.

\section{R-parity violating interactions under the $U(2)_q \times U(2)_\ell$ flavour symmetry}
\label{sec:RPV_and_U2}

The most general renormalizable, R-parity odd superpotential consistent with the gauge symmetry and field content of the MSSM is~\cite{Barbier:2004ez}

\begin{equation}
W_{\text{RPV}} = \mu_i H_u L_i + \frac{1}{2} \lambda_{ijk} L_i L_j E_k^c + \lambda'_{ijk} L_i Q_j D_k^c + \frac{1}{2} \lambda''_{ijk} U_i^c U_j^c D_k^c,
\label{eq:RPVW}
\end{equation}
where there is a summation over the generation indices $i,j,k = 1,2,3$, and summation over gauge indices is understood. One has for example, $L_i L_j E_k^c = \left( \epsilon_{ab} L_i^a L_j^b \right) E_k^c = \left(N_i E_j - E_i N_j\right) E_k^c$, where $a,b=1,2$ are $SU(2)_L$ indices.  \par
Gauge invariance enforces antisymmetry of the $\lambda_{ijk}$ couplings with respect to their first two indices,

\begin{equation}
\lambda_{ijk}=-\lambda_{jik}.
\end{equation}
Gauge invariance also enforces antisymmetry of the $\lambda''_{ijk}$ couplings with respect to their last two indices,

\begin{equation}
\lambda''_{ijk}=-\lambda''_{ikj}.
\end{equation}
Eq. (\ref{eq:RPVW}) involves 48 parameters: 3 dimensionful parameters $\mu_i$ mixing the charged lepton and down-type Higgs superfields, and 45 dimensionless Yukawa-type couplings divided into 9 $\lambda_{ijk}$ and 27 $\lambda'_{ijk}$ couplings which break lepton-number conservation, and 9 $\lambda''_{ijk}$ couplings which break baryon-number conservation. For the sake of simplicity, we shall assume that these couplings are real numbers. \par
The standard motivation for R-parity is that it leads to conserved baryon number and thus ensures proton stability. Nevertheless, according to modern theoretical developments~\cite{Brust:2011tb}, if the MSSM is an effective theory, rapid proton decay can also be induced by higher-dimensional, non-renormalizable operators suppressed by a scale lower than $10 ~\rm TeV$. Consequently, one needs to impose a stand-alone baryon-number conservation symmetry rather than R-parity. In this case, the trilinear terms controlled by the couplings $\lambda$ and $\lambda'$ can be present, while $\lambda''=0$. \par
Regarding the leptonic terms with the coupling $\lambda$, they may seem at first irrelevant in the context of the B-physics anomalies, but as we discuss in Section \ref{sec:constraints} they can contribute to the $b \to s ll$ decays at one-loop level.

\subsection{Tree-level four-fermion operators}
\label{sec:RPV_4fermion}

Expanded in standard four-component Dirac notation, the trilinear interaction terms associated with the $\lambda$ and $\lambda'$ couplings of the RPV superpotential (\ref{eq:RPVW}) read, respectively,
\begin{align}
\label{eq:trilinearlambda}
\mathcal{L}_\lambda&= -\frac{1}{2} \lambda_{ijk} \left(\tilde{\nu}_{Li} \bar{\ell}_{Rk} \ell_{Lj} + \tilde{\ell}_{Lj} \bar{\ell}_{Rk} \nu_{Li} + \tilde{\ell}_{Rk}^* \bar{\nu}_{Ri}^c \ell_{Lj} - (i \leftrightarrow j) \right) + \text{h.c.} \\
\label{eq:trilinearlambdaprime}
\mathcal{L}_{\lambda'}&= - \lambda'_{ijk} \left(\tilde{\nu}_{Li} \bar{d}_{Rk} d_{Lj} + \tilde{d}_{Lj} \bar{d}_{Rk} \nu_{Li} + \tilde{d}_{Rk}^* \bar{\nu}_{Ri}^c d_{Lj} - \tilde{\ell}_{Li} \bar{d}_{Rk} u_{Lj} - \tilde{u}_{Lj} \bar{d}_{Rk} \ell_{Li} - \tilde{d}_{Rk}^* \bar{\ell}_{Ri}^c u_{Lj} \right) + \text{h.c.}
\end{align}
Exchanging Sparticles, one obtains the following four-fermion operators at tree-level:

\begin{align}
\label{eq:four-fermion}
{\cal L}_{eff} &= {\lambda'_{ijk}\lambda^{'*}_{i'j'k}\over 2 m^2_{\tilde
d^k_R}} \Big[ \bar \nu^{i'}_L \gamma^\mu \nu^i_L \bar d^{j'}_L
\gamma_\mu d_L^j + \bar \ell^{i'}_L \gamma^\mu \ell^i_L 
(\bar u_L V_{CKM})^{j'}\gamma_\mu (V_{\text{CKM}}^{ \dagger}u_L)^j \notag \\
&\hspace{2.2cm} -\bar \nu^{i'}_L \gamma^\mu \ell^i_L \bar d^{j'}_L \gamma_\mu (V_{\text{CKM}}^{ \dagger}u_L)^j 
-\bar \ell^{i'}_L \gamma^\mu \nu^i_L (\bar u_LV_{\text{CKM}})^{j'} \gamma_\mu
d^j_L \Big] \notag \\
&-{\lambda'_{ijk}\lambda^{'*}_{i'jk'}\over 2m^2_{\tilde d^j_L} }
\bar \nu^{i'}_L \gamma^\mu \nu^i_L \bar d^k_R \gamma_\mu d^{k'}_R
-{\lambda'_{ijk}\lambda^{'*}_{i'jk'}\over 2m^2_{\tilde u^j_L}}
\bar \ell^{i'}_L \gamma^\mu \ell^i_L \bar d^k_R \gamma_\mu
d^{k'}_R  \notag \\
&-{\lambda'_{ijk}\lambda^{'*}_{ij'k'}\over 2m^2_{\tilde \ell^i_L}}
 (\bar u_{L\beta} V_{\text{CKM}})^{j'} \gamma^\mu (V_{\text{CKM}}^{ \dagger}u_{L\alpha})^j \bar d^k_{R\alpha}
\gamma_\mu d^{k'}_{R\beta} - {\lambda'_{ijk}\lambda^{'*}_{ij'k'}\over 2m^2_{\tilde \nu^i_L}}
\bar d^{j'}_{L\beta} \gamma^\mu d^j_{L\alpha} \bar d^k_{R\alpha}
\gamma_\mu d^{k'}_{R\beta}.
\end{align}
We can simplify the low-energy spectrum by assuming a large mass splitting between the light third generation and the much heavier first two generations of Sfermions and Sleptons, which can be considered as effectively not supersymmetrized. We note, that this simplification is equally well-motivated in theory~\cite{Brust:2011tb}~\cite{Papucci:2011wy}. In fact, most models of spontaneous SUSY breaking predict a significantly lighter third generation at the electroweak scale due to the large RG effects proportional to the top Yukawa coupling $y_t$~\cite{Martin:1997ns}. What is more, as recently shown~\cite{Altmannshofer:2017poe}, one of the prominent attributes of SUSY, namely the gauge coupling unification is still preserved despite the decoupling of the first two generations and even in presence of RPV interactions.

\subsection{Flavour Structure}
\label{sec:U2}

The flavour group we are considering is $\mathcal{G}_f=U(2)_q \times U(2)_\ell$, under which the superfields transform as:
	\begin{align}
	&(Q_1, Q_2) \sim (2,1),  \ \ \ \  Q_3 \sim (1,1), \notag  \\ 
	&(U_1, U_2) \sim (2,1),  \ \ \ \ \  U_3 \sim (1,1), \notag   \\
  &(D_1, D_2) \sim (2,1),  \ \ \ \  D_3 \sim (1,1), \\ 
  &(L_1, L_2) \sim (1,2),  \ \ \ \ \ L_3 \sim (1,1), \notag   \\  
  &(E_1, E_2) \sim (1,2),  \ \ \ \ \ E_3 \sim (1,1). \notag   
\end{align}
The Higgs fields $H_{u(d)}$ are pure $\mathcal{G}_f$ singlets. \par
We introduce additional heavier `flavon' fields, which are charged under the flavour symmetry~\cite{Barbieri:1995uv}; in particular a doublet $\phi_i^a$, a triplet (a 2-index symmetric tensor) $S_i^{ab}$ and a singlet (a 2-index antisymmetric tensor) $A_i^{ab}$, where $i=q,\ell$. The flavour groups $U(2)_i$, are then broken by the vacuum expectation values (VEVs) of these flavon fields, such that,
	\begin{equation}
\left\langle \phi_i^a\right\rangle= M (0 \ \ \epsilon_i)^T, \ \ \ \ \left\langle S_i^{ab}\right\rangle= M \left(\begin{matrix}0 & 0 \\ 0 & \epsilon_{iS} \end{matrix}\right), \ \ \ \ \left\langle A_i^{ab}\right\rangle=M \epsilon'_i \epsilon^{ab},
	\end{equation}
where $M$ is the cut-off of the effective theory. \par
A step-wise breaking is achieved, if we naturally assume $\epsilon \gg \epsilon'$,
	\begin{equation}
U(2)_q \times U(2)_\ell \xrightarrow{\epsilon_q, \epsilon_{qS}, \epsilon_\ell, \epsilon_{\ell S}} U(1)_q \times U(1)_\ell \xrightarrow{\epsilon'_q, \epsilon'_\ell} \text{nothing}
	\end{equation}
The mass matrices of the charged leptons and the down quarks assume the following form:
\begin{equation}
\mathcal{M}_l = \left(\begin{matrix}0 & \epsilon'_\ell & 0 \\ -\epsilon'_\ell & \epsilon_{\ell S} & \epsilon_{\ell} \\ 0 & 0 & 1 \end{matrix}\right) y_{\tau} v_d, \ \ \ \ \ \mathcal{M}_d = \left(\begin{matrix}0 & \epsilon'_q & 0 \\ -\epsilon'_q & \epsilon_{qS} & \epsilon_q \\ 0 & 0 & 1 \end{matrix}\right) y_b v_d,
\end{equation}
where $v_d = v/\sqrt{2} \simeq 174~\rm GeV$ (where $v$ is the SM VEV). Choosing $\epsilon_q = \epsilon_{qS}  \approx m_{s}/m_{b} \simeq 0.025$ and $\epsilon'_q \approx \epsilon_q \sqrt{m_d/m_s}\simeq 0.005$, the usual quark mass hierarchy is successfully reproduced. In the lepton sector, as it will become clear from the phenomenological analysis of Section \ref{sec:constraints}, if one hopes to generate any considerable contribution to the $b \to s \ell\bar{\ell}$ processes the effective coupling to muons must remain unsuppressed. This translates into a strong breaking of $U(2)_\ell$ to $U(1)_\ell$ by the doublet flavon VEV, i.e. $\epsilon_\ell=1$. The correct lepton mass matrix is then reproduced for $\epsilon_{\ell S} \simeq 0.06$ and $\epsilon'_\ell \simeq 0.004$. \par
The RPV bilinear and trilinear terms in the superpotential can be obtained by appropriately contracting the superfields appearing in Eq. (\ref{eq:RPVW}) with the flavons. The order of magnitude of the RPV couplings is the governed by $\epsilon$ and $\epsilon'$, \par
\begin{itemize}
	\item $\lambda_{ijk}$ couplings:
\begin{align}
\label{eq:lambda}
&(121),(131),(133) \sim 0; \ \ (123),(132),(231) \sim \epsilon'_\ell; \ \ (232) \sim \epsilon_{\ell S}; \ \ (122) \sim \epsilon_{\ell} \epsilon'_\ell;  \notag \\
&(233) \sim \epsilon_{\ell};
\end{align}
	\item $\lambda'_{ijk}$ couplings:
\begin{align}
\label{eq:lambdaprime}
&(1jk)',(211)',(231)',(213)',(311)',(331)',(313)' \sim 0; \ \ (221)',(212)' \sim \epsilon_\ell \epsilon'_q;  \notag \\
&(321)',(312)' \sim \epsilon'_q; \ \ (222)',(223)',(232)' \sim \epsilon_\ell \epsilon_q;  \notag \\ 
&(322)',(323)',(332)' \sim \epsilon_q; \ \ (233)' \sim \epsilon_\ell; \ \ (333)' \sim 1.
\end{align}	
\end{itemize}

Generic RPV couplings $\lambda_{ijk}$ (or $\lambda'_{ijk}$) can then be decomposed as products of $\mathcal{O}(1)$ parameters $c_{ijk}$ (or $c_{ijk}'$) and the respective $\epsilon$ and $\epsilon'$ suppression factors.

\section{Constraints from low-energy observables} 
\label{sec:constraints}

In this Section, we analyse the main experimental constraints on the RPV interactions. The processes of interest are the ones that are affected by contributions of the $\mathcal{O}(1)$ couplings $\lambda_{323}$, $\lambda'_{233}$ and $\lambda'_{333}$ or at least by the $\epsilon$-suppressed couplings $\lambda'_{223}$, $\lambda'_{232}$, $\lambda'_{323}$ and $\lambda'_{332}$. These include the $R_{D^{(*)}}$ and $R_{K^{(*)}}$ ratios, the $B_s - \bar{B_s}$ mixing, the $B \to K^{(*)} \nu \bar{\nu} $ and $B \to \tau \bar{\nu}$ decays, the RGE effects in $\tau \to \ell \nu \bar{\nu}$ and the $Z$ coupling modification for the relevant $\lambda'$ couplings and the $\tau \to \ell \nu \bar{\nu}$ decays for the only relevant $\lambda$ coupling. We have explicitly checked that further processes that have been discussed in the bibliography, e.g. the decays $B \to \pi \nu \bar{\nu}$, $B \to \rho \nu \bar{\nu}$, $B \to K \tau \mu$, $B \to X_s \ell^+ \ell^-$, $B \to X_s \gamma$, $B \to \tau \bar{\nu}$, $B \to \tau^+ \tau^-$, $D \to \tau  \bar{\nu}$, $D \to \mu^+ \mu^-$, $\tau \to K \nu$, $\tau \to \pi \nu$, $\tau \to \mu \gamma$ and $\tau \to 3\mu$, do not lead to any relevant constraints in our setup. Of course, all processes involving only the first two generations do not receive any contributions at all. \par

\subsection{$B \to D^{(*)} \tau \bar{\nu}$} 
\label{sec:BtoDtaunu}
 
Adding to the SM the RPV contribution generated by the respective operators in Eq. (\ref{eq:four-fermion}), one obtains the effective Lagrangian describing $b \to c$ semi-leptonic decays at tree-level,

\begin{equation}
\label{eq:btoc}
\mathcal{L}(b\to c\ell\bar{\nu}_{\ell}) = -\frac{4G_F}{\sqrt{2}}V_{cb}(\delta_{ii'}+\Delta^c_{ii'})\bar \ell^{i'}_L \gamma^\mu \nu^i_L \bar c_L \gamma_\mu b_L ,
\end{equation}
where
\begin{equation}
\Delta^c_{ii'} = \sum_{j'=s,b} \frac{\sqrt{2}}{4G_F}\frac{\lambda'_{i33}\lambda'_{i'j'3}}{ 2 m^2_{\tilde
b_R}}\frac{V_{cj'}}{V_{cb}} .
\end{equation}
The ratios (\ref{eq:RDS})-(\ref{eq:RD}) can then be easily written as:
\begin{equation}
\label{eq:RD_theory}
r_{D^{(*)}} = \frac{R_{D^{(*)}}}{R_{D^{(*)}}^{\text{SM}}}  = \frac{\left|1+\Delta^c_{33}\right|^2+\left|\Delta^c_{23}\right|^2}{\frac{1}{2}\left(1+\left|1+\Delta^c_{22}\right|^2+\left|\Delta^c_{32}\right|^2\right)}.
\end{equation}
From the above definition and the weighted average of the $R_D$ and $R_{D^*}$ central values and errors, 
\begin{equation}
r_{D^{(*)}}^{\text{exp}} = 1.227 \pm 0.061,
\end{equation}
one observes that the enhancement of $R_{D^{(*)}}$ implies rather large $\lambda'_{333}$ coupling and $\lambda'_{323} \lambda'_{333}$ coupling combination (which is also enhanced by $V_{cs}/ V_{cb}$), while at the same time $\lambda'_{233}$ and $\lambda'_{223} \lambda'_{233}$ must be kept small. 

\subsection{$B \to K^{(*)} \ell \bar{\ell}$} 
\label{sec:BtoKll}

Instead of using the ratios (\ref{eq:RK}) and (\ref{eq:RKS}), we will instead regard the NP modification of the Wilson Coefficients $C_9$, $C_9'$, $C_{10}$ and $C_{10}'$ defined as:
\begin{align}
\label{eq:btos}
\mathcal{L}(b\to s\ell \bar \ell) = \frac{4G_F}{\sqrt{2}}\frac{\alpha_e}{4 \pi}V_{tb}V_{tb}^* &\left[(C_9^{\ell}+\delta C_9^{\ell})\bar \ell^{i'} \gamma^\mu \ell^i \bar s_L \gamma_\mu b_L + (C_{10}^{\ell}+\delta C_{10}^{\ell})\bar \ell^{i'} \gamma^\mu \gamma_5 \ell^i \bar s_L \gamma_\mu b_L\right. \notag \\
&\left.+({C'}_9^{\ell}+\delta {C'}_9^{\ell})\bar \ell^{i'} \gamma^\mu \ell^i \bar s_R \gamma_\mu b_R + ({C'}_{10}^{\ell}+\delta {C'}_{10}^{\ell})\bar \ell^{i'} \gamma^\mu \gamma_5 \ell^i \bar s_R \gamma_\mu b_R\right],
\end{align}

\begin{figure}[t]
\centering
\includegraphics[angle=0,width=7.3cm]{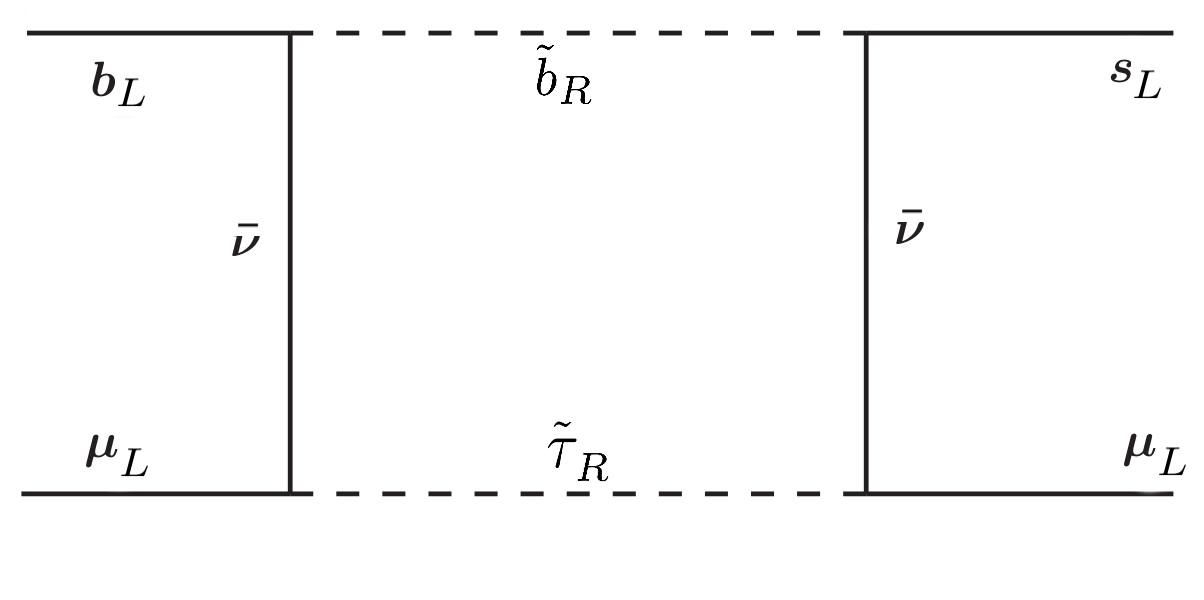}
\caption{The box diagram for $b \to s \mu^+ \mu^-$ transitions with the combination of both leptonic and leptoquark-like couplings.
}
\label{fig:box}
\end{figure}
\FloatBarrier

In our framework, the operator $\bar \mu^{i'}_L \gamma^\mu \mu^i_L \bar d^k_R \gamma_\mu d^{k'}_R$ is generated at tree-level (see Eq. (\ref{eq:four-fermion})) and the operator $\bar \mu^{i'}_L \gamma^\mu \mu^i_L \bar d^k_L \gamma_\mu d^{k'}_L$ at one-loop level~\cite{Das:2017kfo} and thus they give rise to the correlations $\delta C_{9}^{\mu} = -\delta C_{10}^{\mu}$ and $\delta {C'}_{9}^{\mu} = -\delta {C'}_{10}^{\mu}$. \par 
The analysis in~\cite{Altmannshofer:2017yso} provides the best-fit values for different scenarios with NP in one individual Wilson Coefficient at a time and the rest of them SM-like. The relevant results are
\begin{align}
&\left[\delta C_9^{\mu}\right]_{\text{exp}} = -\left[\delta C_{10}^{\mu}\right]_{\text{exp}} = -0.63 \pm 0.17, \\ 
&\left[\delta {C'}_9^{\mu}\right]_{\text{exp}} = -0.05 \pm 0.26, \label{eq:deltaC9prime} \\ 
&\left[\delta {C'}_{10}^{\mu}\right]_{\text{exp}} = -0.03 \pm 0.24 \label{eq:deltaC10prime}.
\end{align}
In the following, the upper index $\mu$ is implied for the Wilson Coefficients. \par
Let us examine first the tree-level case,
\begin{equation}
\delta C'_9 = -\delta {C'}_{10}^{\mu} = \frac{\pi \sqrt{2}}{G_F \alpha_e}\frac{1}{V_{tb}V_{ts}^*}\frac{\lambda'_{232} \lambda'_{233}}{4 m_{\tilde{t}_L}^2}.
\end{equation}
According to Eq. (\ref{eq:lambdaprime}), we expect $\lambda'_{232} \lambda'_{233} \sim \epsilon_\ell^2 \epsilon_q \approx 0.025$ and hence $\delta C'_9 \approx (8\times 10^6 ~\rm GeV^2) /  m_{\tilde{t}_L}^2$. Unless we introduce a significant deviation from the flavour symmetry expectation for the order of magnitude of $\lambda'_{232} \lambda'_{233}$, the left-handed Stop has to be very heavy $m_{\tilde{t}_L}^2 \gg 10 ~\rm TeV$ in order to satisfy the experimental results in (\ref{eq:deltaC9prime}) and (\ref{eq:deltaC10prime}). We are choosing the latter and generalize the result for all left-handed Sparticles. Indeed, we confirm later on that the limit of decoupled left-handed Sparticles is favored in our context. We also note, that for this reason the solutions of~\cite{Das:2017kfo} and~\cite{Earl:2018snx} for the $R_K^{(*)}$ anomaly are not applicable. \par
Next, the NP effect at one-loop level is
\begin{align}
\label{eq:C9}
\delta C_{9}^{\mu} = -\delta C_{10}^{\mu} &= \frac{m_t^2}{16 \pi \alpha}\frac{(\lambda'_{233})^2}{m_{\tilde{b}_R}^2}-\frac{\lambda'_{i23}\lambda'_{i33}\lambda'_{2j3}\lambda'_{2j3}}{64 \sqrt{2} G_F \pi V_{tb}V_{ts}^* \alpha m_{\tilde{b}_R}^2}-\frac{\lambda'_{i23}\lambda'_{i33}\lambda'_{2j3}\lambda'_{2j3}}{64 \sqrt{2} G_F \pi V_{tb}V_{ts}^* \alpha}\frac{\log \left(m_{\tilde{t}_L}^2/m_{\tilde{\nu}_L}^2\right)}{m_{\tilde{t}_L}^2-m_{\tilde{\nu}_L}^2} \notag \\
&-\frac{\lambda'_{323}\lambda'_{333}(\lambda_{323})^2}{64 \sqrt{2} G_F \pi V_{tb}V_{ts}^* \alpha}\frac{\log \left(m_{\tilde{b}_R}^2/m_{\tilde{\tau}_R}^2\right)}{m_{\tilde{b}_R}^2-m_{\tilde{\tau}_R}^2}.
\end{align}
The first term corresponds to box diagrams with a $W$ boson and $\tilde{b}_R$ in the loop, the second to box diagrams with two $\tilde{b}_R$ and the third to a box diagram with $\tilde{\nu}_L$ and $\tilde{t}_L$. The final term is new in our analysis and arises from the diagram in Fig. \ref{fig:box}. The first two terms are unable to explain the $R_K^{(*)}$ anomaly due to severe constraints from $B \to K^{(*)} \nu \bar{\nu}$~\cite{Deshpand:2016cpw} and the third term is suppressed in our framework due to very heavy left-handed Sparticle mediators. It is left to see at what degree can the final term alleviate the tensions, once we have taken the rest of the constraints into consideration, especially those that concern the leptonic currents (see Sec. \ref{sec:tautolnunu}). 

\subsection{$B \to K^{(*)} \nu \bar{\nu}$} 
\label{sec:BtoKnunu}

We define the ratio:
\begin{equation}
R_{B \to K^{(*)} \nu \bar{\nu}} = \frac{ \mathcal B(B \to K^{(*)} \nu \bar{\nu})_{NP}}{ \mathcal B(B \to K^{(*)} \nu \bar{\nu})_{SM} }.
\end{equation}
The relevant effective Lagrangian at tree-level reads
\begin{equation}
\mathcal{L}(b\to s\nu\bar{\nu}) = -\frac{4G_F}{\sqrt{2}} \frac{\alpha}{2 \pi  s_W^2} X_t V_{ts}^* V_{tb} \left( \delta_{ii'}+\frac{\chi_{ii'}}{X_t V_{ts}^* V_{tb}} \right) \bar \nu^{i'}_L \gamma^\mu \nu^i_L \bar s_L \gamma_\mu b_L ,
\end{equation}
where 
\begin{equation}
\chi_{ii'} = -\frac{\pi s_W^2}{\sqrt{2}G_F \alpha}\left( \frac{\lambda'_{i33}\lambda'_{i'23}}{2 m^2_{\tilde
b_R}} \right),
\end{equation}
and $X_t = 1.469 \pm 0.017$ is a SM loop function involving the top quark~\cite{Brod:2010hi}. In principle, the term $-{\lambda'_{ijk}\lambda^{'*}_{i'jk'}\over 2m^2_{\tilde d^j_L} }
\bar \nu^{i'}_L \gamma^\mu \nu^i_L \bar d^k_R \gamma_\mu d^{k'}_R$ in Eq. (\ref{eq:four-fermion}) can also generate contributions to the decay, but since we are considering the limit of decoupled left-handed Sparticles, they become irrelevant. \par
From to the above Equation, we get~\cite{Deshpand:2016cpw}
\begin{equation}
R_{B \to K^{(*)} \nu \bar{\nu}} = \sum_{i=e,\mu,\tau} \frac{1}{3} \left| 1+ \frac{\chi_{ii'}}{X_t V_{ts}^* V_{tb}} \right|^2 + \sum_{i\neq j} \frac{1}{3} \left| \frac{\chi_{ii'}}{X_t V_{ts}^* V_{tb}} \right|^2.
\end{equation}
At $95\%$ confidence level, this ratio is strictly bounded from above~\cite{Buttazzo:2017ixm}
\begin{equation}
R_{B \to K^{(*)} \nu \bar{\nu}} < 5.2
\end{equation}
and consequently, the combinations $\lambda'_{233} \lambda'_{223}$, $\lambda'_{323} \lambda'_{333}$, $\lambda'_{333} \lambda'_{223}$ and $\lambda'_{233} \lambda'_{323}$ are strongly constrained.

\subsection{$B \to \tau \bar{\nu}$} 
\label{sec:Btotaunu}

The $B \to \tau \bar{\nu}$ decay is induced by a $b \to u$ semi-leptonic transition. Analogously to Eq. (\ref{eq:btoc}), we find
\begin{equation}
\label{eq:btou}
\mathcal{L}(b\to u\ell\bar{\nu}_{\ell}) = -\frac{4G_F}{\sqrt{2}}V_{ub}(\delta_{ii'}+\Delta^u_{ii'})\bar \ell^{i'}_L \gamma^\mu \nu^i_L \bar u_L \gamma_\mu b_L ,
\end{equation}
where
\begin{equation}
\Delta^u_{ii'} = \sum_{j'=s,b} \frac{\sqrt{2}}{4G_F}\frac{\lambda'_{i33}\lambda'_{i'j'3}}{ 2 m^2_{\tilde
b_R}}\frac{V_{uj'}}{V_{ub}} .
\end{equation}
We build the ratio:
\begin{equation}
R_{B \to \tau \bar{\nu}} = \frac{ \mathcal B(B \to \tau \bar{\nu})_{NP}}{ \mathcal B(B \to \tau \bar{\nu})_{SM} } = \left|1+\Delta^u_{33}\right|^2
\end{equation}
and then the comparison of the SM prediction~\cite{Altmannshofer:2017poe}
\begin{equation}
\mathcal B(B \to \tau \bar{\nu})_{\text{SM}} = (0.947 \pm 0.182) \times 10^{-4}
\end{equation}
with the experimental average~\cite{Amhis:2016xyh}
\begin{equation}
\mathcal B(B \to \tau \bar{\nu})_{\text{exp}} = (1.06 \pm 0.19) \times 10^{-4}
\end{equation}
yields relevant bounds on $\lambda'_{333}$ and $\lambda'_{323} \lambda'_{333}$. 

\subsection{$B_s - \bar{B_s}$ mixing} 
\label{sec:B-Bbar}

The effective Hamiltonian relevant to $\Delta B = 2$ processes is

\begin{equation}
\mathcal{L}(\Delta B = 2) = - C_{1 i}^{VLL} (\bar{b}_L \gamma^{\mu} d_\ell^i)^2 - C_{2 i}^{LR} (\bar{b}_R^{\alpha} d_L^{i \alpha}) (\bar{b}_L^{\beta} d_R^{i \beta}) + 2 C_{1 i}^{LR} (\bar{b}_R^{\alpha} d_L^{i \beta}) (\bar{b}_L^{\beta} d_R^{i \alpha}) ,
\end{equation}
where we have used the notation of Eq. (2.1) in~\cite{Buras:2001ra} and expressed the Wilson Coefficient of the operator $Q_{1 s}^{LR}$, accordingly, by performing a Fierz transformation to the operator $Q_5$ in the so-called `SUSY basis'~\cite{Gabbiani:1996hi}. This operator arises at tree-level by mediation of a left-handed Sneutrino $\bar{\nu}_L$, while the SM-like operator $(\bar{b}_L \gamma^{\mu} d_\ell^i)^2$ and the operator $(\bar{b}_R^{\alpha} d_L^{i \beta}) (\bar{b}_L^{\beta} d_R^{i \alpha})$ appear first only at one-loop level by various box diagrams~\cite{Wang:2010vv}. We find
\begin{equation}
C_{2 s}^{LR} = \frac{\lambda'_{332}\lambda'_{323}}{2 m_{\bar{\nu}_L}^2},
\end{equation}
which in the limit of decoupled left-handed Sparticles vanishes. Further, if we likewise choose to neglect the box diagrams with left-handed Sparticle mediators, we find
\begin{align}
C_{1 s}^{VLL} &= \frac{\lambda'_{i23}\lambda'_{j33}\lambda'_{j23}\lambda'_{i33}}{128 \pi^2 m^2_{\tilde
b_R}}, \\
-2C_{1 s}^{LR} &= -\frac{\lambda'_{i23}\lambda'_{j33}\lambda'_{i32}\lambda'_{j33}}{64 \pi^2 m^2_{\tilde
b_R}}.
\end{align}
The Wilson Coefficients of the rest of the operators do not receive any RPV contributions and finally, we may write the Eqs. (7.25), (7.27) in~\cite{Buras:2001ra} as:
\begin{align}
\Delta M_{B_s} = 2\left|\bra{B^0}\mathcal{H}_{\text{eff}}^{\Delta B = 2} \ket{\bar{B}^0}\right| = \frac{2 m_B F_B^2}{3} \left|P_1^{VLL}C_{1 s}^{VLL}+P_1^{LR}C_{1 s}^{LR}+P_2^{LR}C_{2 s}^{LR}\right|,
\end{align}
where
\begin{equation}
P_1^{VLL}=0.84, \ \ P_1^{LR}=-1.62  \ \ \text{and} \ \ P_2^{LR}=2.46.
\end{equation}
Finally, using the experimental bounds~\cite{Amhis:2014hma} 
\begin{equation}
\Delta M_{B_s}^{\text{exp}} = (1.1689 \pm 0.0014) \times 10^{-11} ~\rm GeV
\end{equation}
and the SM prediction~\cite{Bazavov:2016nty}
\begin{equation}
\Delta M_{B_s}^{\text{SM}} = (1.2903 \pm 0.1316) \times 10^{-11} ~\rm GeV,
\end{equation}
bounds are set for all contributing $\lambda'$ coupling combinations. 

\subsection{$Z \to \ell \bar{\ell '}$ coupling} 
\label{sec:tautolnunu}
As investigated in~\cite{Feruglio:2016gvd}~\cite{Feruglio:2017rjo}, the leptonic $Z$ coupling is modified via one-loop diagrams. In our setup, the $Z$ boson decays to a $t\bar{t}$ pair, which is in turn connected by a virtual $\tilde{b}_R$ and eventually turns into a dilepton pair $\ell \bar{\ell '}$. One defines the following ratios of vector and axial-vector couplings $v_\ell$ and $a_\ell$:
\begin{align}
&\frac{v_{\tau}}{v_{e}}=1-\frac{2\delta g_{\ell L}^{33}}{1-4s_W^2}, \\
&\frac{a_{\tau}}{a_{e}}=1-2\delta g_{\ell L}^{33}.
\end{align}
In our context\footnote{We match the effective Lagrangian of Eq. (19) with ours (\ref{eq:four-fermion}) by setting $C_1=\frac{1}{2}$, $C_3=-\frac{1}{2}$ and \\ $\lambda_{i'i}^e \lambda_{j'j}^u=\lambda'_{ijk} \lambda^{'}_{i'j'k}$.} and keeping only the term proportional to the top Yukawa, the Eq. (30) in~\cite{Feruglio:2017rjo} becomes:
\begin{equation}
\delta g_{\ell L}^{33} \simeq \frac{3y_t^2}{32\sqrt{2}G_F \pi^2 }\frac{(\lambda'_{333})^2}{m_{\tilde{b}_R}^2} \left(\log \left(\frac{\Lambda}{m_Z}\right) - 0.612 \right).
\end{equation}
We denote the NP scale to be roughly $\Lambda \simeq m_{\tilde{b}_R} \approx 1 ~\rm TeV$. \par
A comparison with the measured values~\cite{Patrignani:2016xqp},
\begin{equation}
\left[\frac{v_{\tau}}{v_{e}}\right]_{\text{exp}}=0.959 \pm 0.029 \ \ \text{and} \ \ \left[\frac{a_{\tau}}{a_{e}}\right]_{\text{exp}}=1.0019 \pm 0.0015,
\end{equation} 
yields bounds on the coupling $\lambda'_{333}$. \par
We note here, that the $W$ coupling is also modified, but the constraints given by LFU violating $\tau$ decays on the same couplings are more stringent (see Sec. \ref{sec:tautolnunu}).

\subsection{$\tau \to \ell  \nu \bar{\nu}$} 
\label{sec:tautolnunu}

The purely leptonic operators resulting from the trilinear couplings in (\ref{eq:trilinearlambda}), affect the decays $\tau \to e \nu \bar{\nu}$ and $\tau \to \mu \nu \bar{\nu}$ at tree-level through the exchange of a third generation Slepton $\tilde{\tau}_R$~\cite{Kao:2009fg}. Additionally, the RGE effects driven by the top Yukawa $y_t$ interactions contribute also to the decay width~(\cite{Feruglio:2017rjo}, via one-loop diagrams involving the leptoquark-like interactions in (\ref{eq:trilinearlambdaprime}). The NP effects are probed by the ratio:
\begin{equation}
\label{eq:Rtauldef}
R_{\tau}^{\tau/\ell } = \frac{ \mathcal B(\tau \to \ell  \nu \bar{\nu})_{\text{exp}} / \mathcal B(\tau \to \ell  \nu \bar{\nu})_{\text{SM}}}{ \mathcal B(\mu \to e \nu \bar{\nu})_{\text{exp}} / \mathcal B(\mu \to e \nu \bar{\nu})_{\text{SM}}} = \sum_{ij}\left|\delta_{i3}\delta_{\ell j}+\frac{1}{2} r_{ij3}+(C_L^{\tau/\ell })_{ij}\right|^2,
\end{equation}
where
\begin{equation}
r_{ij3}=\frac{\sqrt{2}}{4G_F}\frac{(\lambda_{ij3})^2}{2m_{\tilde{\tau}_R}^2}
\end{equation}
and
\begin{equation}
(C_L^{\tau/\ell })_{ij} = \frac{y_t^2}{32\sqrt{2} G_F\pi^2 }\frac{\lambda'_{333}}{m_{\tilde{b}_R}}\left[3\delta_{ij}\lambda'_{\ell 33}-3(\delta_{lj}\lambda'_{i33}+\delta_{i3}\lambda'_{\ell j3})\right]\log{\left(\frac{\Lambda}{m_Z}\right)}.
\end{equation}
At leading order, Eq. (\ref{eq:Rtauldef}) becomes:
\begin{equation}
\label{eq:Rtaul}
R_{\tau}^{\tau/\ell } \simeq 1 + \frac{\sqrt{2}}{4G_F}\frac{(\lambda_{323})^2}{m_{\tilde{\tau}_R}^2} -\frac{3y_t^2}{16\sqrt{2} G_F\pi^2 }\frac{(\lambda'_{333})^2}{m_{\tilde{b}_R}^2}\log{\left(\frac{\Lambda}{m_Z}\right)}.
\end{equation}
Even though the experimental bounds are very stringent~\cite{Pich:2013lsa}, i.e. 
\begin{equation}
\left[R_{\tau}^{\tau/\mu}\right]_{\text{exp}} = 1.0022 \pm 0.0030 \ \ \text{and} \ \ \left[R_{\tau}^{\tau/e}\right]_{\text{exp}}=1.0060 \pm 0.0030, 
\end{equation}
with apporpriate fine-tuning of the couplings and the masses, one could still recover a non-negligible leptonic current interaction.

\section{Numerical fit and discussion} 
\label{sec:fit}

After the decoupling of left-handed Sparticle related contributions, the low-energy observables discussed above depend solely on the RPV couplings $\lambda_{323}$, $\lambda'_{223}$, $\lambda'_{232}$, $\lambda'_{233}$, $\lambda'_{323}$, $\lambda'_{332}$ and $\lambda'_{333}$ and on the masses $m_{\tilde{b}_R}$ and $m_{\tilde{\tau}_R}$. \par
We  have  performed  a  combined  fit  of  these parameters using as input the experimental data reported in Sec. \ref{sec:constraints} and various SM parameters~\cite{Patrignani:2016xqp}. The following, conservative, lower bounds for the right-handed Sparticles are imposed: $m_{\tilde{b}_R} > 800 ~\rm GeV$ and $ m_{\tilde{\tau}_R} > 400 ~\rm GeV$~\cite{Patrignani:2016xqp}. We have also assumed that the $\mathcal{O}(1)$ parameters $c_{ijk}$ (or $c_{ijk}'$), which together with the flavour suppression factors constitute the RPV couplings, are restricted by the unitarity bounds of $\sqrt{4 \pi}$. For simplicity, we  have assumed Gaussian errors for all the observables. The preferred region of the model parameters $x$ has been determined minimizing the $\chi^2$ distribution:
\begin{equation}
\chi^2(x) = \sum_i \left(\frac{\mathcal O_i(x) - \mu_i}{\sigma_i}\right)^2, 
\end{equation}
where $\mu_i$ and $\sigma_i$ are the central values and the $1\sigma$ uncertainties of the measured values $\left[\mathcal O_i\right]_{\text{exp}}$, respectively. \par
The best-fit points for the $c_{ijk}$ (or $c_{ijk}'$) parameters, as well as the flavour parametric scaling and the total value of the relevant RPV couplings are listed in the following Table:

\begin{table}[htb]
\centering
\begin{tabu}{|c|[1pt]c|c|c|}
\hline
RPV couplings    & $c_{ijk}$ (or $c_{ijk}'$) best-fit point & Parametric Scaling         & Total value \\ \tabucline[1pt]{-} 
$\lambda_{323}$  & 4              & $\epsilon_q$            & 0.12        \\ \hline
$\lambda'_{223}$ & -0.2           & $\epsilon_q \epsilon_\ell$ & -0.006      \\ \hline
$\lambda'_{232}$ & -0.1           & $\epsilon_q \epsilon_\ell$ & -0.003      \\ \hline
$\lambda'_{233}$ & 0.1            & $\epsilon_\ell$            & 0.1         \\ \hline
$\lambda'_{323}$ & 4              & $\epsilon_q$               & 0.12        \\ \hline
$\lambda'_{332}$ & 0.1            & $\epsilon_q$               & 0.003       \\ \hline
$\lambda'_{333}$ & 0.66           & 1                          & 0.66        \\ \hline
\end{tabu}
\caption{The best-fit points along with the flavour suppression factors for each of the relevant RPV couplings are reported.}
\label{tbl:bfp}
\end{table}
The best-fit points for the masses are $m_{\tilde{b}_R} = 900 ~\rm GeV$ and $m_{\tilde{\tau}_R} = 9 ~\rm TeV$. \par
\begin{figure}[t!]
\centering
\includegraphics[width=0.475\textwidth]{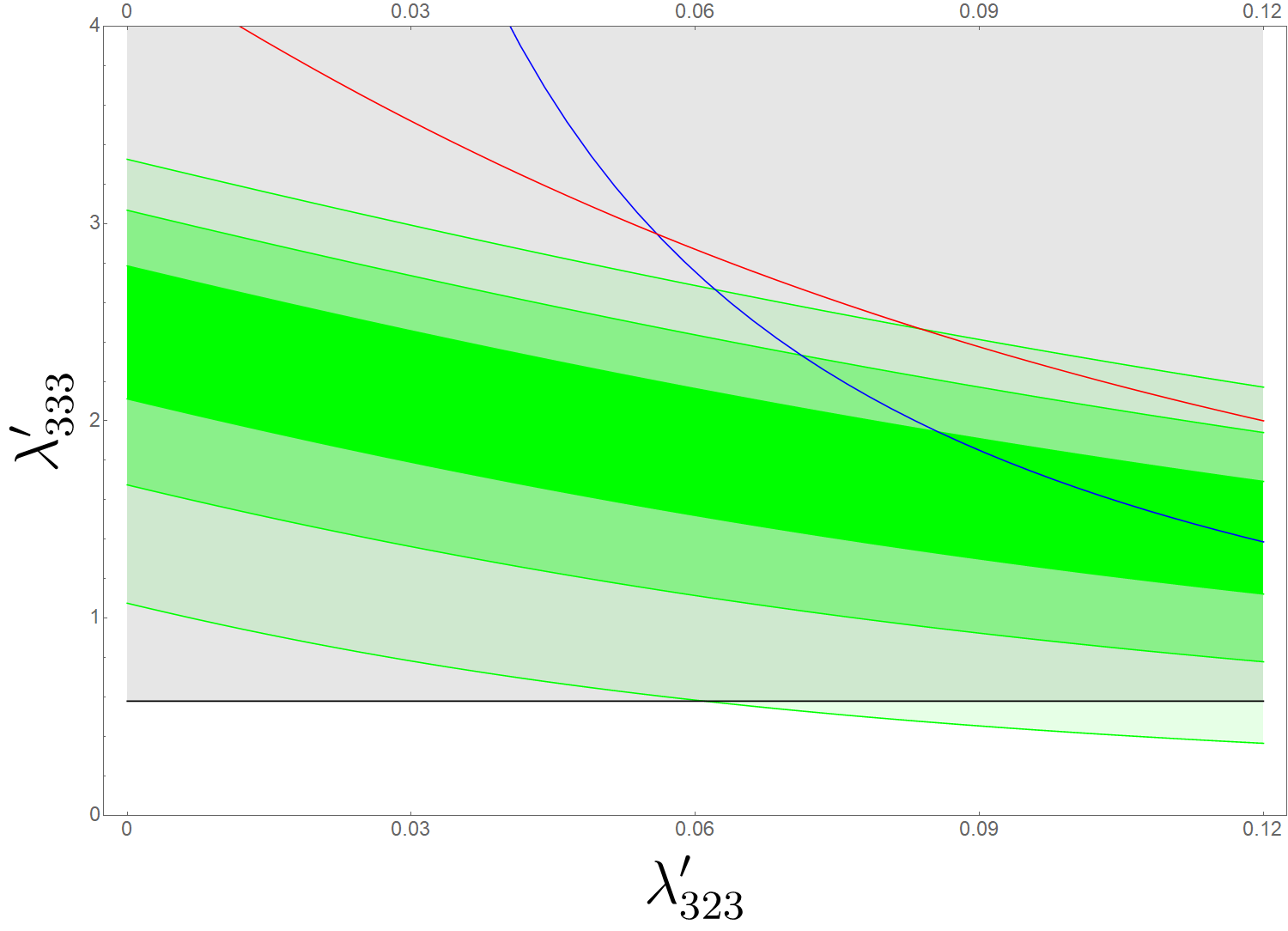} ~~~~~ 
\includegraphics[width=0.475\textwidth]{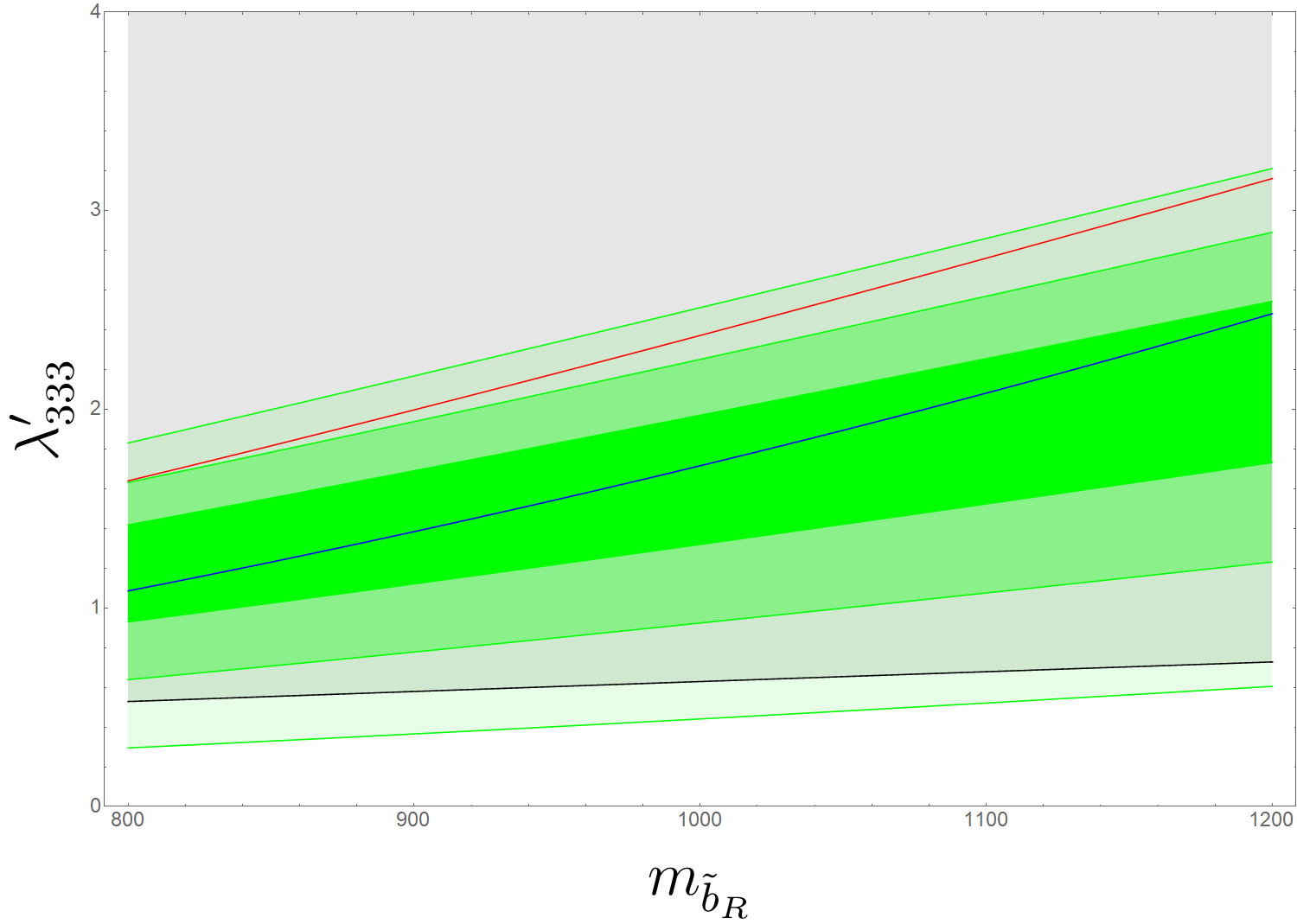} \\[5pt]
\includegraphics[width=0.475\textwidth]{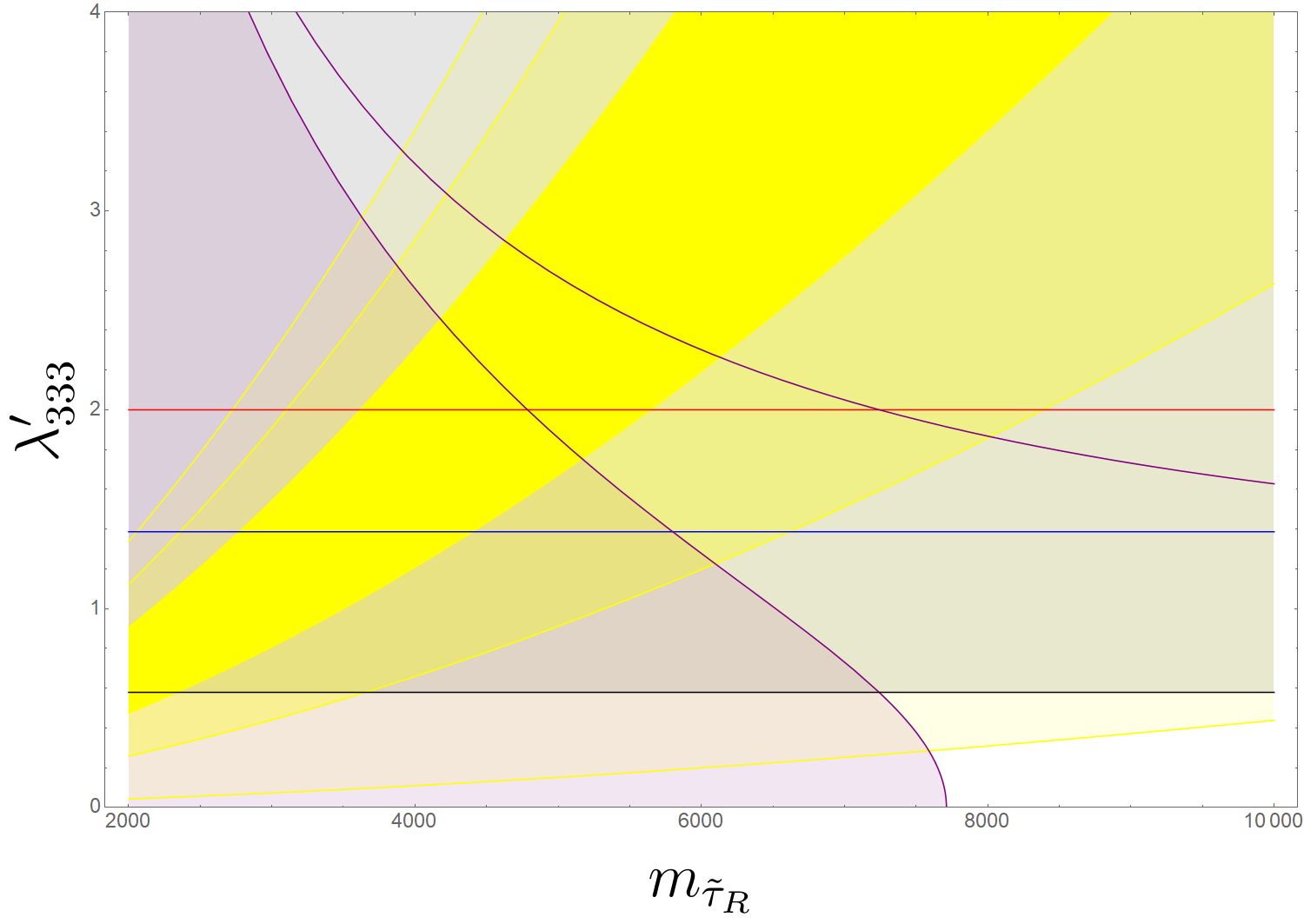} ~~~~~
\includegraphics[width=0.175\textwidth]{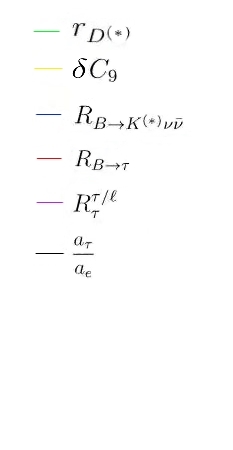} ~~~~~
\caption{RPV parameter space accommodating $r_{D^{(*)}}$ (green) and $\delta C_9$ (yellow) at $1 \sigma$, $2 \sigma$ and $3 \sigma$ around the best-fit point. The constraints originating from the following observables: $R_{B \to K^{(*)} \nu \bar{\nu}}$ (blue), $R_{B \to \tau \bar{\nu}}$ (red), $R_{\tau}^{\tau/\ell }$: (purple), $\frac{a_{\tau}}{a_{e}}$ (black) are shown. The parameter space above the contours is excluded at $2 \sigma$.}
\label{fig:2Dplanes}
\end{figure}
\begin{figure}[t]
\centering
\includegraphics[width=0.55\textwidth]{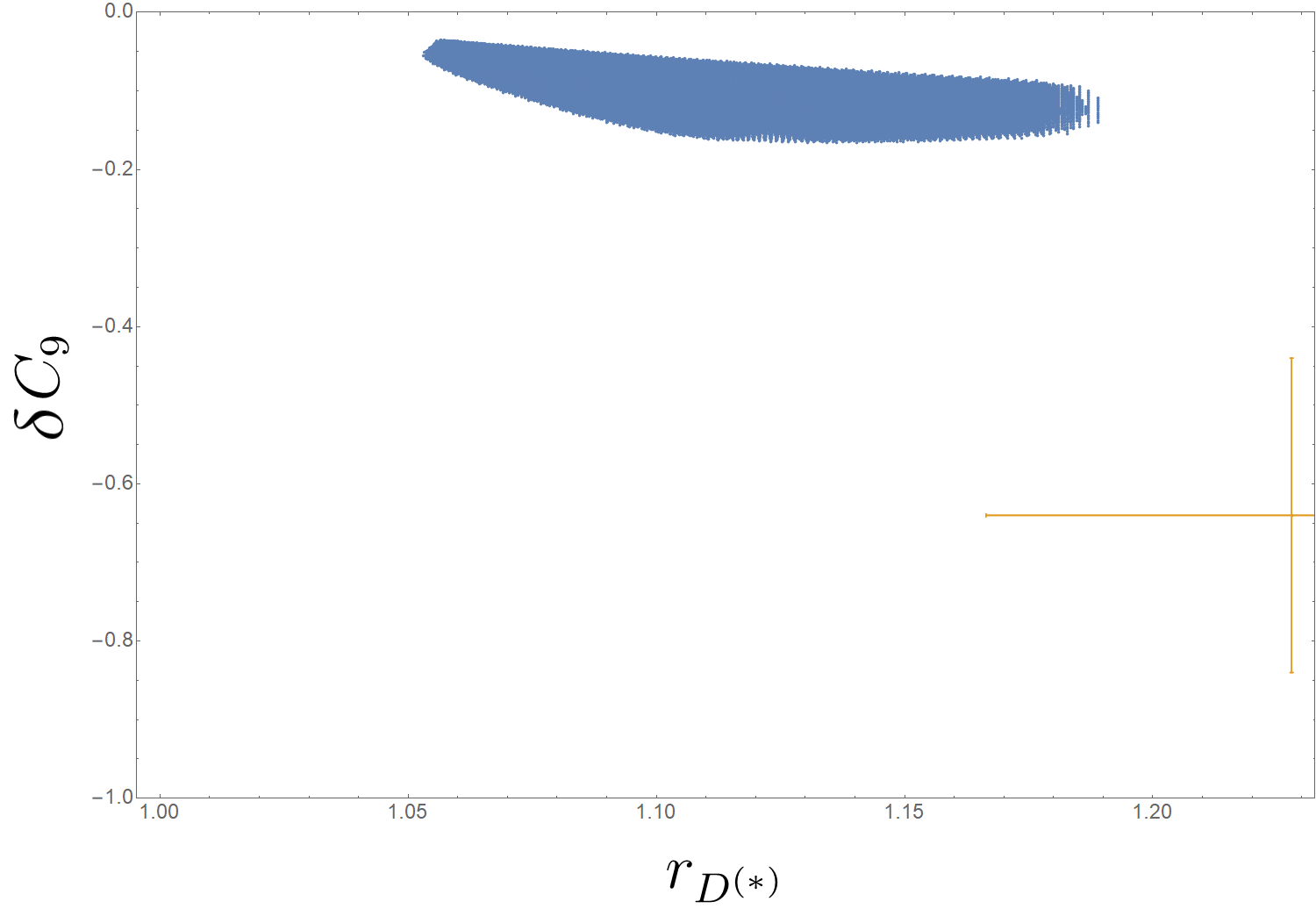}
\caption{Model prediction in the $\delta C_9$ vs. $r_{D^{(*)}}$ plane for the $\Delta \chi^2 < 2.3$ ($1\sigma$) region around the best-fit point. The $1\sigma$ experimental data are shown by the cross. The SM prediction coincides with the origin of the axes (top-left).}  
\label{fig:fit}
\end{figure}

The improvement of the best-fit point of the total $\chi^2$ with respect to the SM limit is $\chi^2 (x_{\text{SM}} ) − \chi^2 (x_{\text{BF}} ) = 9.43$. In Table \ref{tbl:chi2components}, we show the contributions of the individual summands in the $\chi^2$.
\begin{table}[htb]
\centering
\begin{tabu}{|c|[1pt]c|c|}
\hline
Observables    & $\chi_i^2 (x_{\text{SM}})$ & $\chi_i^2 (x_{\text{BF}})$ \\ \tabucline[1pt]{-} 
$r_{D^{(*)}}$  &  13.71     &  5.27     \\ \hline
$\delta C_9$  &  10.24     &  8.09     \\ \hline
$R_{B \to \tau \bar{\nu}}$  &  0.34     &  0.86     \\ \hline
$\Delta M_{B_s}$  &  2.34     &  0.12     \\ \hline
$\frac{v_{\tau}}{v_{e}}$  &  1.99     &  1.24     \\ \hline
$\frac{a_{\tau}}{a_{e}}$  &  1.60     &  3.82     \\ \hline
$R_{\tau}^{\tau/\ell }$  &  0.53     &  0.86     \\ \hline
\end{tabu}
\caption{The individual $\chi^2$ components evaluated at the best-fit point and compared with the respective SM limits.}
\label{tbl:chi2components}
\end{table}
\newpage
In Fig. \ref{fig:2Dplanes} we show  the $68\%$ CL and $95\%$ CL regions of the $r_{D^{(*)}}$ and $\delta C_9$ observables in the $(\lambda'_{333},\lambda'_{323})$, $(\lambda'_{333},m_{\tilde{b}_R})$, $(\lambda'_{323},m_{\tilde{b}_R})$ and $(\lambda'_{333},m_{\tilde{\tau}_R})$ planes, after having fixed the rest of the parameters according to Table \ref{tbl:bfp}. The constraints of the other, relevant low-energy observables are presented as exclusion contours at $2\sigma$. The degree of consistency of the best fit-region with the anomalies is illustrated in Fig. \ref{fig:fit}, where we show the values of the two observables $r_{D^{(*)}}$ and $\delta C_9$ within the $1\sigma$ preferred region ($\Delta \chi^2 < 2.3$). \par
First and foremost, we observe that our choice of flavor symmetry points towards an alleviation of tensions in the charged-current anomalies similar to the results in~\cite{Altmannshofer:2017poe}\footnote{We confirm that our fit is at least as good as the one intended to solve the charged-current anomalies with generic RPV couplings up to the small term proportional to $\lambda'_{313}$. The bounds from the $Z$ coupling modification taken in the current paper are also more stringent that the ones in~\cite{Altmannshofer:2017poe}, because we constraint directly the axial-vector coupling $a_\ell$, instead of the left-handed coupling $g_{\ell L}^{33}$. Moreover, there is an additional factor of $1/2$ suppressing the NP contribution, which is not taken into account in Eq. (9) in~\cite{Altmannshofer:2017poe}, as opposed to Eq. (4) in~\cite{Deshpand:2016cpw}.}. As a matter of fact, it was commonly expected that new phenomena would not show up in tree-level processes, where the effect has to be comparable with the SM value, but rather in FCNC transitions where one has to simply compete against a SM loop-suppressed contribution. In this regard, it is interesting that the flavour structure favors an improvement of the tree-level anomalies rather than the loop-induced ones. Nevertheless, our model prediction can only exhibit a $1\sigma$ agreement with the present central value of $R_{D^{(*)}}$ in case of a more than $2\sigma$ reduction of the ratio ${a_{\tau}}/{a_{e}}$. This is, indeed, the main obstacle for obtaining a larger $\lambda'_{333}$, required for a perfect fit. Other bounds, including those from the $B_s - \bar{B_s}$ mixing and the $B \to \tau \bar{\nu}$ decays, are found to be less significant. \par
Regarding $R_{K^{(*)}}$, most of the parameter space is excluded due to the bounds from the $R_{\tau}^{\tau/\mu}$ observable . In principle, an appropriate canceling of the second and the third term in (\ref{eq:Rtaul}) can lift this constraint, but considering the previous bounds on $\lambda'_{333}$, we can achieve at most a $\sim 10\%$ cancellation at $2\sigma$ deviation from the central value. This still allows for a smaller mass for $m_{\tilde{\tau}_R}$ and as a result, a slightly better fit for $\delta C_9$.
\newpage

\FloatBarrier

In total, we see that, by involving the leptonic RPV interactions, we can `cure' approximately $30 \%$ of the $b \to s\ell^+\ell^-$ anomalies, when all other constraints are allowed to deviate by maximum $2\sigma$ from their experimental average. This is still an improvement over previous attempts to accommodate both anomalies within the RPV framework. Please note, that the fit yields very small $\lambda'_{233}$ and $\lambda'_{233}$ couplings and thus both the second term in (\ref{eq:C9}) and the NP muonic contribution in the denominator of (\ref{eq:RD_theory}) are negligible. This implies that there is no clash between the anomalies themselves, but rather the performance of the fit is limited by the other low-energy constraints. \par
On a final note, we would like to briefly comment on three, relevant issues. First, it is well-known, that RPV interactions that attempt to address LFU violating effects can also generate neutrino masses at one-loop level~\cite{Barbier:2004ez},
\begin{equation}
\Delta M_{\nu,ij}^{\lambda'} \simeq \frac{3}{8\pi^2}\frac{m_b^2(A_b-\mu \tan \beta)}{m_{\tilde{b}_R}^2}\lambda'_{i33}\lambda'_{j33}.
\end{equation}
The typical way-out is to postulate a direct cancellation between the trilinear coupling $A_b$ and the term $\mu \tan \beta$, which is though hard to justify theoretically. Other unrelated NP contributions to neutrino masses, e.g. the standard see-saw mechanism with heavy right-handed neutrinos, could be the cause of the suppression. We stress, here that the possibility of generating a canceling due to loop effects induced by the leptonic trilinear RPV interactions is ruled out due to the direct, tree-level bounds from $\tau$ decays on the $\lambda$ couplings. \par
The next comment refers to the resulting SUSY mass spectrum. According to the fit, the mass of the right-handed superpartners spans the {1 - 10 ~\rm TeV} range, with the $\tilde{\tau}_R$ being significantly heavier than the $\tilde{b}_R$. In fact, only the mass scale of $\tilde{b}_R$ appears to be within the reach of high-$p_T$ searches in the near future. An even more striking assumption of our setup is the complete decoupling of the left-handed superparters. A theoretical motivation for this scenario lies in the theory of gauge-mediated supersymmetry breaking. As it turns out, LHC observable Squarks can only contribute enough quantum corrections to lift the Higgs mass if the left-handed $SU(2)_L$ doublets are much heavier than the singlets~\cite{Shao:2012je}.\par
Last but not least, as shown in~\cite{Crivellin:2016ejn}, flavour symmetries similar to the one employed here, when gauged and broken at the $~\rm TeV$ scale, can provide a natural mechanism for generating additional LFU violating contributions to $b \to s\ell^+\ell^-$ transitions. A further discussion of this scenario, that would require an enlargement of the field content of the theory, is beyond the scope of this paper.  

\section{Conclusions}

In this paper we have studied the consistency of the $R_{D^{(*)}}$ and $R_{K^{(*)}}$ anomalies with all relevant low-energy observables, in the context of RPV interactions controlled by a $\mathcal{G}_f=U(2)_q \times U(2)_\ell$ flavour symmetry. This particular scenario favors a viable solution of the charged-current anomaly, at least as good as the generic, effective RPV-SUSY scenario. However, as we have shown, a perfect fit of the anomalous $R_{D^{(*)}}$ observable cannot be achieved without a significant modification of the $Z$ boson coupling, occurring at one-loop level. What is more, upon inclusion of the leptonic, trilinear RPV interactions, we were able to generate, simultaneously, a contribution to the $b \to s\ell^+\ell^-$ transitions, which is also limited by the same $Z$ boson coupling bounds and the tree-level, Lepton Flavor violating $\tau$ decays. All in all, the flavour symmetry for natural values of the free parameters, as summarised in Table \ref{tbl:bfp}, can provide an explanation of the strength of the RPV interactions, required for a better fit of the B-physics anomalies. 

\section*{Acknowledgements}
We thank Gino Isidori and Javier Fuentes-Martin for useful  discussions and comments on the manuscript. This research was supported in part by the Swiss National Science Foundation (SNF) under contract 200021-159720.

\end{document}